\documentclass[epsf,12pt]{article}
\usepackage{graphicx}
\begin{document}
\begin{center}
{\Large\bf \mbox{Glass Transition Temperature and Fractal Dimension}
           \mbox{of Protein Free Energy Landscapes} 
      }
\vskip 1.1cm
{\bf Nelson A. Alves$^{a,}$\footnote{alves@quark.ffclrp.usp.br}
  ~and~ 
Ulrich H.E. Hansmann$^{b,}$\footnote{hansmann@mtu.edu} }\\
\vskip 0.3cm
{$^a$Departamento de F\'{\i}sica e Matem\'atica, FFCLRP \\
     Universidade de S\~ao Paulo. Av. Bandeirantes 3900. CEP 014040-901 \\ 
     Ribeir\~ao Preto, SP, Brazil \\
\vskip 0.3cm
   $^b$Department of Physics, Michigan Technological University \\
     Houghton, MI 49931-1291, USA 
 }
\end{center}
\vspace{0.7cm}
%%%\baselineskip=0.8cm
%%%%%%%%%%%%%%%%%%%%%%%%%%%%%%%%%%%%%%%%%%%%%%%%%%%%%%%%%%%%%%%
\newenvironment{abs}{\begin{quote}}{\end{quote}}
\begin{abs}
\centerline{\bf Abstract}
 The free-energy landscape of two peptides is evaluated at various 
 temperatures
 and an estimate for its fractal dimension at these temperatures calculated. 
 We show that  monitoring this quantity as a function of temperature allows
 to determine the glass transition temperature.
\vskip 0.1cm
{\it Keywords:} Energy landscape, folding funnel, fractal dimension,
                protein folding, generalized-ensemble simulations  
\end{abs}
%%%%%%%%%%%%%%%%%%%%%%%%%%%%%%%%%%%%%%%%%%%%%%%%%%%%%%%%%%%%%%%%%
It is well known that a large class of proteins folds spontaneously
into unique, globular shape  which is determined solely
by the sequence of amino acids (the monomers) \cite{Anf}. 
Folding of a protein  into this three-dimensional structure (in which
it is biologically active) is now often described 
by  energy landscape theory and the funnel concept \cite{OLSW}. In 
this ``new view'' of folding \cite{DC} it is assumed that the 
energy landscape of a protein is rugged, however, unlike for 
a random hetero-polymer,  there is a sufficient overall slope so
that the numerous folding routes converge towards the native structure.
The particulars of the  funnel-like energy landscape determine
the transitions between the different thermodynamic states
\cite{OLSW,DC}.  For instance, a common scenario for folding may be
that first the protein collapses from a random coil to a compact state.
This coil-to-globular transition is characterized by the
collapse transition temperature $T_{\theta}$. In the second stage,
a set of compact structures is explored. The final stage involves a
transition, characterized by the folding temperature $T_f~ (\le T_{\theta})$,
from one of the many local minima in 
the set of compact structures to the native conformation.

However, the essence of the funnel landscape idea is competition between the
tendency towards the folded state and trapping due to ruggedness of
the landscape. It is expected 
that for a good folder the temperature, $T_g$, where glass behavior
sets in, has to be significantly lower than the folding temperature
$T_f$ \cite{Bryngelson87}. 
It follows that it is important to calculate these two temperatures when
examining the folding of a protein in  computer simulations.
While the folding temperature $T_f$ can be easily 
determined by monitoring the changes of a suitable order parameter 
with temperature, the situation is less obvious for the  
glass transition temperature $T_g$ which is normally determined from
the slowing down of folding times with temperature in kinetic 
simulations \cite{OLSW}.  To measure $T_g$ from equilibrium properties
of the protein one can use the
intimate connection between ``roughness'' and fractality.
Especially, we expect that the fractal dimension of  the folding funnel
will increase with increasing roughness of the {\it free} energy 
landscape. We conjecture that the 
glass transition temperature $T_g$ is associated with
a change of the fractal dimension of the folding funnel  and
propose  to measure $T_g$ by calculating the fractal dimension
of the protein free energy landscape as a function of temperature. We remark
that our approach differs from Ref.~\cite{LTEG} where recently for 
some peptides 
the fractal properties of the time series of the
{\it potential} energy were  studied. 
 
While landscape theory and funnel concept were developed from
studies of minimalistic protein models without reference to 
specific proteins, we intend to probe our assumption 
for two distinct peptides.
The first peptide is Met-enkephalin, which 
has the amino acid sequence Tyr-Gly-Gly-Phe-Met. In previous work
\cite{HMO97b,HOO98b} evidence was presented
 that the folding of this peptide can be
described within the funnel concept. Estimators for the
collapse temperature $T_{\theta}=295\pm 30$ K and the folding temperature
$T_f = 230 \pm 30$ K were presented \cite{HMO97b}. As the second 
peptide we choose poly-alanine of chain length $N=20$,
% I REMEMBER WE AGGREED IN SUMMER THAT N=30 DATA WERE NOT RELIABLE, OR? 
which undergoes at $T=508(5)$ K a sharp transition between a completely
ordered helical state and a random (coil) state \cite{OH95a,HO98c}.
Hence, we expect no
finite glass transition temperature  for this polypeptide; and 
the thermal behavior of the fractal dimension should differ 
significantly from that of Met-enkephalin.

Our simulations of both peptides relied on a detailed, realistic,
description of the intramolecular interactions. Such simulations are
known to be notoriously difficult.
This is because at low temperatures simulations based on canonical 
Monte Carlo or molecular dynamics techniques will 
get trapped in one of the  multitude of local minima separated
by high energy barriers, and physical quantities cannot be calculated 
accurately. Only recently, with the development of 
 {\it generalized-ensemble} techniques such as
 multicanonical sampling \cite{MU}
 and simulated tempering \cite{L}, calculation of 
accurate low-temperature thermodynamic  quantities  became feasible 
in protein simulations \cite{HO,OH95a,HO98c}. Hence, the use of one of these 
novel techniques, multicanonical sampling \cite{MU},
 was crucial for our project. 

In a multicanonical algorithm \cite{MU} conformations with energy $E$ are
assigned a weight $ w_{mu} (E)\propto 1/n(E)$, $n(E)$ being the density of
states.  A simulation with this weight generates a random walk in the energy
space  and a large range of energies is sampled. Hence, 
one can use the re-weighting techniques \cite{FS} to calculate the free
energy $G(x)$ as a function of the chosen reaction coordinate $x$
over a wide temperature range by
\begin{equation}
G(x,T) = -k_B T \log [P(x) w_{mu}^{-1}(E(x))\, e^{-E(x)/k_B T}] - C~.
\label{eqol}
\end{equation}
Here, $P(x)$ is the distribution of $x$ as obtained by our multicanonical
simulation and $C$ is chosen so that the lowest value of $G(x)$ is set to zero
for each temperature.  Unlike in a canonical simulation the weight 
$ w_{mu}$ is not a priori known, and estimators have to be calculated
using the procedures described in Refs.~\cite{HO,HO96g}.

The main problem in characterizing the roughness of the 
high-dimensional folding funnel of a protein is the choice of
an appropriate reaction coordinate. 
We choose   for Met-enkephalin 
the overlap with the (known) ground state, $O$,
defined by \cite{HMO97b}
\begin{equation}
O = 1 -\frac{1}{90~n_F} \sum_{i=1}^{n_F} |\alpha_i- \alpha_i^{(GS)}|~,
\label{eqop}
\end{equation}
where $\alpha_i^{(GS)}$  stand for
the $n_F$ dihedral angles of the ground state conformation. 
Similar, we chose as order parameter for poly-alanine  the helicity
\begin{equation}
q = \frac{\tilde{n}_H}{N-2}~,
\label{helicity}
\end{equation}
which allows us to distinguish between helical and coil configurations
of poly-alanine.
Here $\tilde{n}_H$ is the number of helical residues in a conformation,
without counting the first and last residues which can move freely and
will not be part of a helical segment. 

The fractal dimension of the free energy landscape can be calculated from 
different definitions, and different definitions can yield different 
information about the graph under study \cite{Falconer}.
 From a theoretical point of view the Hausdorff-Besicovitch 
definition \cite{Falconer,Feder} is the
proper one to characterize the geometrical complexity, 
however,  it is very difficult to evaluate numerically.
More widely used techniques to obtain a dimension of an arbitrary set
are  box-counting (and its generalized version) \cite{Falconer,Feder,Block} 
and the method introduced by Higuchi \cite{Higuchi}, which we used for
our analysis.

To define a fractal dimension, 
Higuchi considers a finite set of observations $X(j),\,
 j=1, 2, ...,\, N$
taken at a regular interval $k$ in the reaction coordinate,
 and evaluates the length  $L_m(k)$ of the corresponding
graph for different interval length $k$
obtained from sequences
\begin{equation}
 X_k^m  : X(m), X(m+k), X(m+2k), ..., X(m+[\frac{N-m}{k}]k)\,, 
                                                   \label{eq:Hig0}
\end{equation}
where $m=1,2,...,k$ and 
$[\frac{N-m}{k}]$ denotes the integer part of $(N-m)/k$.
 The length of the graph is calculated as
\begin{equation}
 L_m(k)  =  
 \frac{N-1}{k}\, 
\sum_{i=1}^{[\frac{N-m}{k}]} \frac{|X(m+ik) - X(m+(i-1)k)|}
                           {k\, [\frac{N-m}{k}]}\,.
\end{equation}
If the behavior of the graph has fractal characteristics over the
available range $k$ then
\begin{equation}
              <L(k)>\, \propto\, k^{-d}~,        \label{eq:Hig2}
\end{equation}
where $d$ is the fractal dimension and $<L(k)>$ is 
the average value over $k$ partial lengths of the graph.

We start now presenting our results 
which rely on 2,000,000 sweeps 
for both Met-enkphalin and  poly-alanine.  
The potential energy function $E_{tot}$ that we used is given
by the sum of
electrostatic term $E_C$, Lennard-Jones term $E_{LJ}$, and
hydrogen-bond term $E_{hb}$ for all pairs of atoms in the peptide
together with the torsion term $E_{tors}$ for all torsion angles.
The parameters for the energy function were adopted from
ECEPP/2 \cite{EC}   (as implemented
in the KONF90 program \cite{Konf}).
We further fix the peptide bond angles $\omega$
 to their common value $180^{\circ}$,  and do 
 not  explicitly include the interaction
of the peptide with the solvent and set the dielectric constant $\epsilon$
equal to 2.

 In Fig.~1 we show the free energy
landscape of Met-enkephalin as a function of the overlap with the ground
state for $T=230$ K. The funnel towards the ground state is clearly visible
in this plot. In previous work \cite{HOO98b} it was found that at this
temperature no long-living traps exist and therefore  the funnel
is relatively smooth.
 In Fig.~2 we show the corresponding logarithm of
the averaged curve length $<L(k)>$ over the interval length $k$,
for this temperature.
 The straight line corresponds to the least-square fit to the
linear model obtained from Eq. (\ref{eq:Hig2}).
 The error bars are the standard deviations obtained from 
$k$ sets $\,L_m(k)$ for the all statistics.
 Here we present the fractal dimension obtained from the whole
statistics instead of introducing any binning procedure of our data.
Hence, the errors reported here for the final estimates of $d$ are 
related to the deviation of the linear behavior of
${\rm ln}<L(k)>$ in Eq. (\ref{eq:Hig2}). 

Repeating the above analysis for various
temperatures, we obtain a plot of the fractal dimension as a function
of temperature which is displayed for Met-enkephalin in Fig.~3. Various
distinct regions can be observed in this graph. For the high temperature
region the fractal dimension seems to be constant and only slightly deviating
from a one-dimensional graph: we find $d \approx 1.15$. 
In Refs.~\cite{HMO97b,HOO98b} it was shown that  this temperature region is
dominated by extended coil structures with little resemblance to the
ground state. Hence the energy landcape is a rapidly increasing function
of the overlap, with the minimum at small values of that order parameter.
With decreasing
temperature the fractal dimension of the free energy landscape increases
till it reaches a local maximum of $d_{1}=1.33\pm 0.05$ for $T=280\pm 40$ K  
(the quoted uncertainty in the temperature  is an upper estimate and 
given by the range of temperatures 
for which the measured fractal dimension  $d$ lies within the errorbars 
of $d_1$). This temperature 
seems to correspond to $T_{\theta} = 295\pm 20$ K, the collapse temperature
found in earlier work \cite{HMO97b} for Met-enkephalin. At $T_{\theta}$
both extended coil structures and an ensemble of collapsed structures
can exist, and the free energy landscape reflects the large fluctuations
at this temperature. In Ref.~\cite{HOO98b}
it was shown that this temperature a funnel-like structure of the landscape
starts to appear, which becomes clearly visible at the folding temperature
$T_f=230\pm  30$ K. We find in our plot of the fractal dimension no indication
for this folding transition, presumably because that temperature seems to
be within the same peak. Instead we observe that the fractal dimension 
decreases again with further decreasing temperature.  
 This is consistent with our previous results \cite{HMO97b} and indicates
that as the temperature decreases the  ground state structure becomes 
more and more favored in the ensemble of compact structures. Actually, the 
folding temperature is  defined by the condition that half of the observed
configurations are ground state - like, and that temperature, 
$T_f=230\pm30$ K
corresponds roughly to the mid point of the fractal dimension plot
between its maximum of $d=1.33\pm 0.05$ for $T=280\pm 40$ K and the
low-temperature minimum of $d=1.25\pm 0.02$ at a temperature 
$T=180\pm 30$ K.
Below that temperature, the fractal dimension increases rapidly again,
indication the onset of glassy behavior and the appearance of long-living
traps. Hence, we identify this temperature  as the glass temperature and
find for Met-enkephalin
\begin{equation}
 T_g = 180 \pm 30 K~.
\end{equation}
This estimate is consistent with $T_g = 160 \pm 30 K$ as determined by
an approximate calculation in Ref.~\cite{H99d}.
 
As mentioned above, it is expected  that for a protein $T_f > T_g$,  
i.e. a good folder can be characterized by the
relation \cite{Bryngelson87}
\begin{equation}
\frac{T_f}{T_g} > 1~.
\end{equation}
The  result for $T_f=230\pm30$ K (as quoted in Ref.~\cite{HMO97b}) 
and our new estimate for the
glass transition temperature $T_g$ lead indeed to
%\begin{equation}
%\frac{T_f}{T_g} =  1.28 > 1~,
%\end{equation}
%where we have used the means to evaluate the above ratio. This clearly
$T_f/T_g = 1.28 > 1$. This value of the ratio clearly
demonstrates that Met-enkephalin is a good folder according to the above 
criterion. 
Our result is  consistent with an alternative characterization of
folding properties. Thirumalai and collaborators \cite{KTh}
have pointed out that  the kinetic accessibility of the native 
conformation can be classified by the parameter
\begin{equation}
\sigma = \frac{T_{\theta} - T_f}{T_{\theta}}~,
\label{sig}
\end{equation}
i.e., the smaller $\sigma$ is, the more easily a protein can fold.
With central values $T_{\theta} = 295$ K and $T_f = 230$ K, taken from
Ref.~\cite{HMO97b}, we have for Met-enkephalin
$\sigma \approx 0.2$   which implies reasonably good folding properties 
according to Ref.\cite{KTh}. Hence, we see that there is a
strong correlation between the folding criterion ($T_f/T_g > 1$) proposed by
Bryngelson and Wolynes \cite{Bryngelson87} and the one by Thirumalai and
co-workers \cite{KTh}.

It is interesting to compare the above graph with the behavior of the fractal
dimension for poly-alanine with its sharp helix-coil transition.
Following the receipt described for  Met-enkephalin, we obtain from an
analysis of free energy landscapes $G(q,T)$ (with $q$ defined in 
Eq.~\ref{helicity}) a plot of the fractal dimension as a function
of temperature for poly-alanine chains of length $N=20$. The
graph is displayed in Fig.~4.  
 
Here we also observe an almost flat curve for high temperatures with a
small value ($d \approx 1$) 
of the fractal dimension of the free energy landscape as
a function of the helicity, which is our order parameter for this system.
The small value for the fractal dimension indicates again the relative
smooth landscape of the peptide in this temperature region which is
dominated by coil structures.
Again the fractal dimension increases with decreasing temperature till a
maximum  of $d=1.6 \pm 0.3$
is reached at $T=510\pm 30$ K for $N=20$. 

This temperature corresponds to
the helix-coil-transition temperature  $T_c=508\pm 5 $ K of 
Ref.~\cite{HO98c}. Below that temperature the fractal
dimension decreases again rapidly, but unlike for Met-enkephalin 
it does not increase again at some lower temperature. 
The rapid decrease in $d$ reflects the
observation that below $T_c$ the system  exist almost exclusively in
a single configuration, namely as a single  extended $\alpha$-helix,
and that the transition between coil and helix states is either 
first-order-like or can be described as a strong second order transition
\cite{HO98c}.
Since the ground state structure is so strongly energetically favored for
poly-alanine, we find no indication for a glass transition at lower 
temperatures.  Such a behavior would be expected for pronounced
transitions of the above type.

Let us summarize our results. We have used generalized-ensemble simulations
to calculate the free-energy landscapes of two peptides as a function of
a suitable reaction coordinate for a large temperature range. We have measured
the fractal dimension of these energy landscapes and studied its thermal
behavior. Our results show that  the fractal dimension $d(T)$ as a function
of temperature is sensitive to thermodynamic transitions in the molecules.
Especially, it is possible to determine estimators for the glass transition
temperature $T_g$ from this quantity.  

%%%%%%%%%%%%%%%%%%%%%%%%%%%%%%%%%%%%%%%%%%%%%%%%%%%%%%%%%%%%%%%%%
\noindent
{\bf Acknowledgements}: \\
U.H.~was visitor at the 
Ribeir\~ao Preto campus of the Universidade de S\~ao Paulo when this work
was performed. He thanks the Departamento de F\'{\i}sica e Matem\'atica
for the kind hospitality extended to him, and FAPESP for a generous
travel grant.  
Financial supports from FAPESP and a Research Excellence
Fund  of the State of Michigan
are gratefully acknowledged.

\newpage
{\Large Figure Captions:}\\
\begin{enumerate}
\item  Free energy of Met-enkephalin 
       as a function of the overlap with the (known)
       ground state O for $T=230$ K. The results  are calculated from a
       generalized-ensemble simulation of 2,000,000 Monte Carlo sweeps.
\item  Linear regression for ln$\,<L(k)>$ (as defined in Eq.~(\ref{eq:Hig2}))  
       for $T=230$ K. 
\item  Fractal dimension of the free-energy landscape of Met-enkephalin
       as function of temperature. 
\item  Fractal dimension of the free-energy landscape of Poly-alanine
       (with chain length $N=20$) as a function of temperature.
\end{enumerate}

\newpage
%FIGURE 1
\begin{figure}[t]
\begin{center}
\begin{minipage}[t]{0.95\textwidth}
\centering
\includegraphics[angle=-90,width=0.72\textwidth]{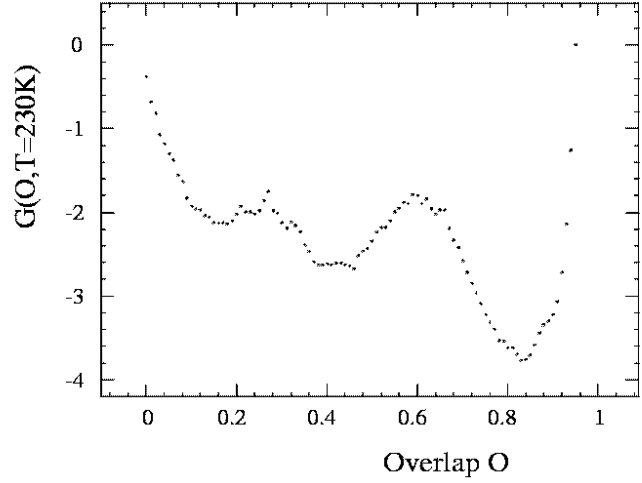}
\renewcommand{\figurename}{FIG.}
\caption{Free energy of Met-enkephalin as a function of the overlap
 with the (known) ground state O for $T=230$ K.
 The results  are calculated from a
       generalized-ensemble simulation of 2,000,000 Monte Carlo sweeps.}
\label{fig1}
\end{minipage}
\end{center}
\end{figure}

\newpage
%FIGURE 2
\begin{figure}[t]
\begin{center}
\begin{minipage}[t]{0.95\textwidth}
\centering
\includegraphics[angle=-90,width=0.72\textwidth]{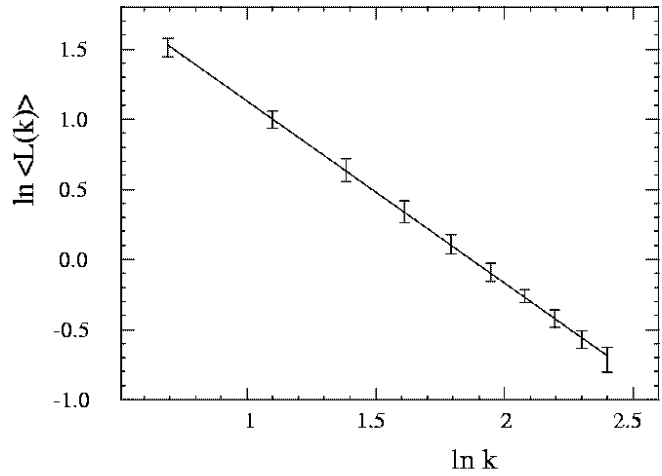}
\renewcommand{\figurename}{FIG.}
\caption{Linear regression for ln$\,<L(k)>$ 
     (as defined in Eq.~(\ref{eq:Hig2})) for $T=230$ K.} 
\label{fig2}
\end{minipage}
\end{center}
\end{figure}

\newpage 
%FIGURE 3
\begin{figure}[t]
\begin{center}
\begin{minipage}[t]{0.95\textwidth}
\centering
\includegraphics[angle=-90,width=0.72\textwidth]{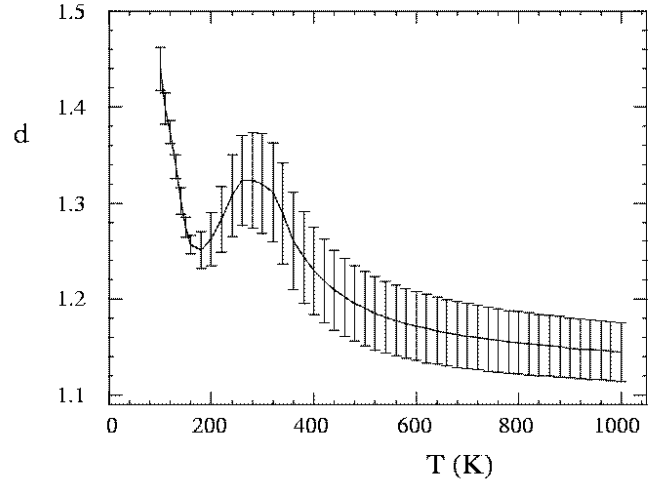}
\renewcommand{\figurename}{FIG.}
\caption{Fractal dimension of the free-energy landscape of Met-enkephalin
       as function of temperature.}
\label{fig3}
\end{minipage}
\end{center}
\end{figure}

\newpage
%FIGURE 4
\begin{figure}[t]
\begin{center}
\begin{minipage}[t]{0.95\textwidth}
\centering
\includegraphics[angle=-90,width=0.72\textwidth]{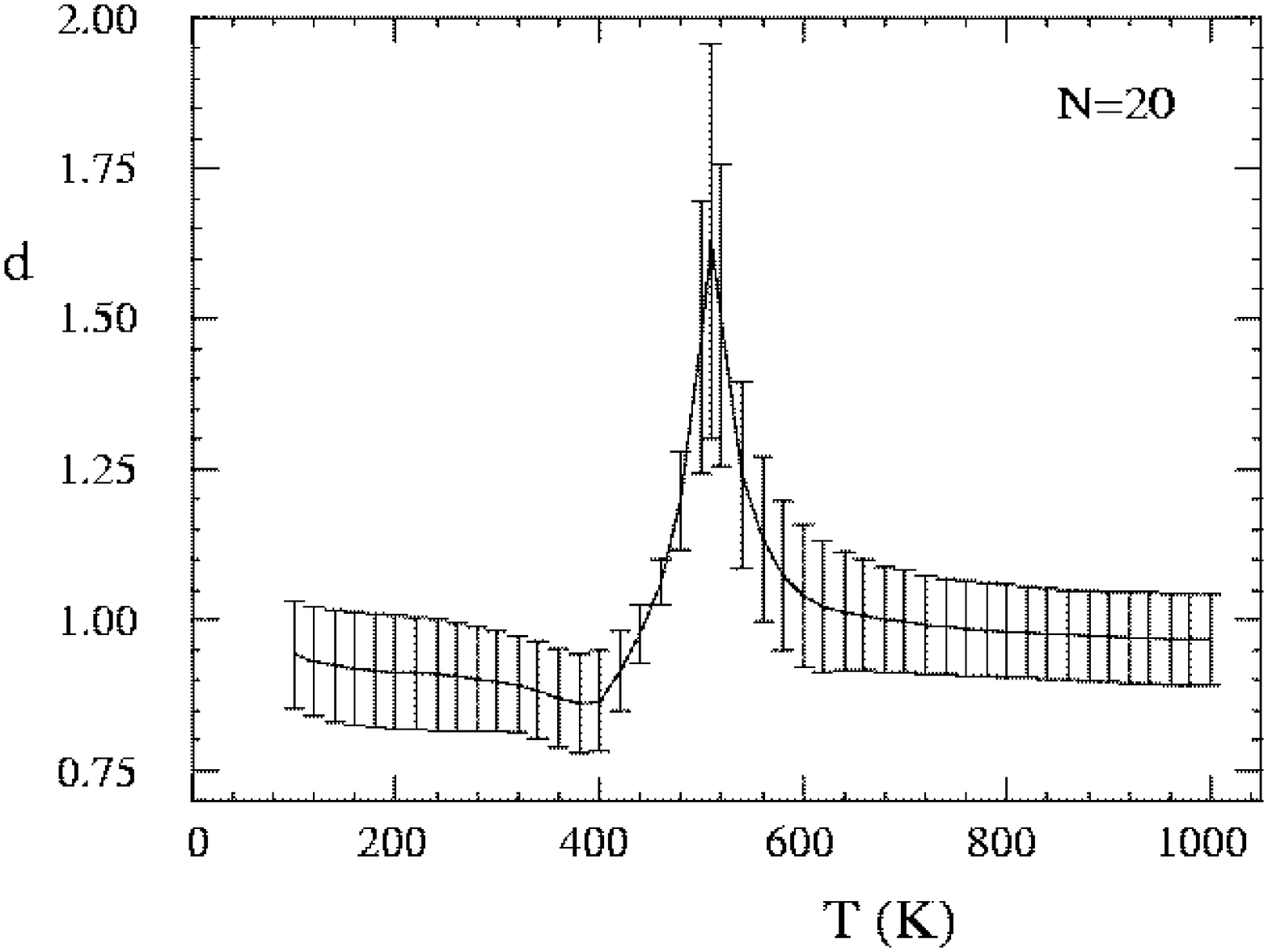}
\renewcommand{\figurename}{FIG.}
\caption{Fractal dimension of the free-energy landscape of Poly-alanine
      (with chain length $N=20$) as a function of temperature.}
\label{fig4}
\end{minipage}
\end{center}
\end{figure}


\begin{thebibliography}{99}
\bibitem{Anf} C.B.~Anfinsen, {\it Science} {\bf 181}, 223 (1973).
\bibitem{OLSW} J.N. Onuchic, Z. Luthey-Schulten, P.G. Wolynes,
        Annual Reviews  in Physical Chemistry {\bf 48}, 545 (1997).
\bibitem{DC} K.A.~Dill and H.S.~Chan, {\it Nature Structural
             Biology} {\bf 4}, 10 (1997).
\bibitem{Bryngelson87} J.D. Bryngelson and P.G. Wolynes, Proc. Natl. Acad.
        Sci. U.S.A. {\bf 84}, 524 (1987);J.N. Onuchic, Z. Luthey-Schulten,
        P.G. Wolynes, Annual Reviews  in Physical Chemistry {\bf 48}, 545
        (1997).
\bibitem{LTEG} D.A.~Lidar, D.~Thirumalai, R.~Elber and R.B.~Gerber,
               {\it Phys.~Rev.~E} {\bf 59}, 2231 (1999).
\bibitem{HMO97b} U.H.E.~Hansmann, M.~Masuya, and Y.~Okamoto,
                 Proc.~Natl.~Acad.~Sci.~U.S.A. {\bf 94}, 10652 (1997).
\bibitem{HOO98b} U.H.E.~Hansmann, Y.~Okamoto, and J.N.~Onuchic,  Proteins
                 {\bf 34}, 472 (1999)
\bibitem{OH95a} Y.~Okamoto and U.H.E.~Hansmann,\  J.~Phys.~Chem.
                {\bf 99}, 11276 (1995).
\bibitem{HO98c} U.H.E.~Hansmann and Y.~Okamoto,  J. Chem.~Phys. {\bf 110},
                1267 (1999); {\bf 111} 1339(E) (1999).
\bibitem{MU} B.A.~Berg and T.~Neuhaus, 
             Phys. Lett. {\bf 267}, 249 (1991).
\bibitem{L}  A.P.~Lyubartsev,~A.A.Martinovski,\ S.V.~Shevkunov, and \
             P.N.\ Vorontsov-Velyaminov,\  J.~Chem.~Phys. {\bf 96},
             1776 (1992);
             E.~Marinari and G.~Parisi, Europhys.~Lett. {\bf 19},
             451 (1992).
\bibitem{HO} U.H.E. Hansmann and Y. Okamoto,  J.~Comp.~Chem.
             {\bf 14}, 1333 (1993).
\bibitem{FS}  A.M. Ferrenberg and R.H. Swendsen, Phys.\ Rev.\ Lett.
              {\bf  61}, 2635 (1988); Phys. Rev. Lett. {\bf 63},
              1658(E) (1989), and references given in the erratum.
\bibitem{HO96g} U.H.E.~Hansmann and Y.~Okamoto, 
                {\it Phys. Rev. E}, {\bf 56}, 2228 (1997).
\bibitem{Falconer} K. Falconer, {\it Fractal Geometry Mathematical
                   Foundatiosn and Applications} (John Wiley \& Sons, 1990).
\bibitem{Feder} J. Feder, {\it Fractals} (Plenum, New York, 1988). 
\bibitem{Block}  A. Block, W. von Bloh, and H.J. Schellnhuber, 
                 {\it Phys. Rev.} {\bf A42}, 1869 (1990), 
                 and references therein.
\bibitem{Higuchi} T. Higuchi, Physica {\bf D31}, 277 (1988);
                  {\bf D46}, 254 (1990). 
\bibitem{EC} M.J. Sippl, G. N{\'e}methy, and H.A. Scheraga,
             {\it J. Phys. Chem.} {\bf 88}, 6231~(1984), 
             and references therein.
\bibitem{Konf} H.~Kawai, Y.~Okamoto, M.~Fukugita, T.~Nakazawa, and
               T.~Kikuchi,  {\it Chem. Lett.} {\bf 1991}, 213 (1991);
               Y.~Okamoto, M.~Fukugita, T.~Nakazawa, and H.~Kawai,
               {\it Protein Engineering} {\bf 4}, 639 (1991).
\bibitem{KTh} D.K.~Klimov and D.~Thirumalai, {\it Phys. Rev. Lett.} {\bf 76},
              4070 (1996).
\bibitem{H99d} U.H.E.~Hansmann, {\it Eur.~Phy.~J.~B} {\bf 12}, 607 (1999).
\end{thebibliography}
\end{document}